\documentclass[aps,pre,reprint]{revtex4-1}
\usepackage [utf8]{inputenc}
\usepackage[T1]{fontenc}
\usepackage{lmodern}
\usepackage{amsmath}
\usepackage{siunitx}
\DeclareSIUnit{\calorie}{cal}
\usepackage{hyperref}
\usepackage{mathrsfs}
\usepackage{multirow}
\usepackage{texdraw}
\usepackage{color}
\usepackage{epstopdf}
\usepackage[font=small,labelfont=bf]{caption}

\usepackage{graphicx}

\usepackage{color}

\begin{document}
\thispagestyle{empty}

% Doc really begins here
\title{Hydrodynamic interactions between solutes in multiparticle collision dynamics}
\author{Vincent Dahirel$^a$, Xudong Zhao$^a$, Baptiste Couet$^a$, Guillaume Bat\^ot$^b$, Marie Jardat$^a$ }
\affiliation{$^a$Sorbonne Universit\'e, CNRS, Physico-chimie des \'electrolytes et nano-syst\`emes interfaciaux, PHENIX, F-75005 Paris, France, \\
$^b$ IFP \'Energies Nouvelles, avenue de Bois Pr\'eau, 92852 Rueil-Malmaison Cedex, France} 
\date{\today}

\begin{abstract}
Multiparticle  collision dynamics (MPCD) enables to simulate fluid dynamics including both hydrodynamics and thermal fluctuations. Its main use concerns complex fluids, where the solvent interacts with concentrated solutes, may they be colloidal particles, polymers or electrolytes. A key difficulty concerns the way one couples the fluid to the solute particles, without losing the key advantages of the MPCD method in term of computational efficiency. In this paper, we investigate the dynamical properties of solutes that are coupled to the fluid within the collision step, {\em i.e.} when local momentum exchange between fluid particles occurs. We quantify how the volume where momentum exchange is performed (the size of the collision cells) constrains the hydrodynamic size of the solute. Moreover, we show that this volume should be taken smaller than the structural size of the solutes. Within these constraints, we find that the hydrodynamic properties of a 1-1 electrolyte solution are similar to the behaviour predicted by the Fuoss-Onsager theory of electrolyte dynamics, and we quantify the limitations of the theory for 2-1 and 2-2 electrolytes.  However, it is also clear that mapping the diffusion time scale to that of a real system cannot be done quantitatively with this methodology.  

\end{abstract}  
\maketitle

\section{Introduction}
The dynamics of solutes in a solution drives many properties, from their thermic or electrical conductivity to the rate of chemical reactions. 
Dynamical properties of solutes depend on hydrodynamic interactions mediated by the solvent. 
These interactions play a role in a very large concentration range. Typically, in electrolytes \cite{bernard1}, they become significant at concentrations larger than $10^{-2}$ mol.L$^{-1}$, and in some case they become screened in very dense solutions, such as polymer melts \cite{AhlrichsPRE2001}. 
Most transport theories of simple solutions, electrolyte solutions, dilute polymers, or colloidal suspensions rely on a description of hydrodynamic interactions derived from the Stokes equation of fluid dynamics at low Reynolds numbers. 
The Fuoss-Onsager theory of electrolyte transport \cite{onsa57,bernard1} and the classical theories of polymer transport \cite{Zimm56,kirkwood,Rotne69,AhlrichsPRE2001} both include hydrodynamics using the Oseen tensor.
In this framework, the velocity of a solute  is influenced by the velocities of surrounding solutes through a tensor depending on the distances between particles. 
The Oseen tensor can be derived by evaluating the effect of a force applied on a fixed point of the fluid described by the Stokes equation.
 The size and shape of the particles have no explicit influence on hydrodynamic interactions within this modelling. 
Finding numerical alternatives to these limited theoretical treatments without resorting to atomistic numerical simulations has given rise to a variety of mesoscopic methods.
 Brownian Dynamics with the Rotne-Prager hydrodynamic tensor \cite{Rotne69} is the simulation technique that is the most natural \cite{ErmakJCP78}, 
as it contains ingredients from classical theoretical treatments. 
This method has been very successful in computing transport coefficients for various systems \cite{JardatJCP99,YamaguchiJCP2011,WajnrybPhysA2004,MeriguetJCP05,RexEPJE2008,JaegerPRE2012,PamiesJPS2007,StoltzJCP2007}
but it has several major pitfalls. First, when the system is too concentrated, 
the Rotne-Prager tensor can be non positive definite for some configurations of the system, 
and the simulation cannot proceed. This is a clear limitation if one
 is interested in crowded media, such as the interior of a biological cell. 
Secondly, when strong attractive interactions between particles exist, 
random displacements can lead the system in regions of phase space that should not be realistically explored, 
which leads to strong instabilities of the simulation. This difficulty can be partially overcome by the use of Metropolis Adjusted Langevin Algorithm (MALA), see {\em e. g.} ref.\cite{MALA}.  
Moreover, it is very difficult to adapt such simulation strategies to concentrated solutions or confined solutions for which the use of the Rotne-Prager tensor is no longer valid\cite{beenaker84}. 

Two popular alternatives to Brownian Dynamics are Dissipative Particle Dynamics \cite{DPD}, and Multi-Particle Collision Dynamics (MPCD)\cite{MalKap,WinklerRev}. 
In the present paper, we focus on the use of the latter to study systems with strong hydrodynamic interactions between solutes. 
In MPCD, an explicit but highly simplified description of the solvent is used, where ballistic motions and local momentum exchanges between 
solvent particles are tuned to reproduce the properties of a fluid at the Navier-Stokes level. 
It is a discrete solvent with the hydrodynamic properties of a continuous solvent, whose hydrodynamic regime can be chosen thanks to a relatively small number of parameters. Solute particles can be embedded in the MPCD solvent bath, and evolve through a classical molecular dynamics algorithm; Hydrodynamic interactions between solutes emerge in this case.
One advantage of MPCD is that it describes hydrodynamics more adequately than Brownian Dynamics with the Rotne-Prager tensor, and that it  can in principle be used for any boundary conditions. Moreover, the algorithm is particularly suited to parallelization as the local momentum exchanges happen in collision cells defined by a grid of fixed size $a_0$. 

There are several ways to couple the solute particles to the MPCD fluid. 
A central repulsive force between the solute and the solvent can be added. It creates a spherical zone around the solute depleted from the solvent \cite{KapralJCP2000, WinklerCPC2005,PaddingPRE06}. 
This scheme leads to an effective slip boundary condition at the surface of the solute. 
Alternative schemes can lead to an effective stick boundary condition \cite{PaddingJPCM05,WhitmerJPCM10,GotzePRE11}. 
In cases where the solute particles are rather small, 
another possibility is  to couple the solute with the solvent during the collision step,
 when the momentum exchange occurs. This coupling scheme is rather crude, 
but it is the most efficient from the computational point of view. 
In the following, we refer to this coupling scheme as collisional coupling (denoted by MPCD-CC in what follows).  
This scheme has been used for instance  to study the dynamics of small polymer chains \cite{MalevanetsEPL2000,Mussawisade2005,FrankEPL08,FrankJCP09}, 
and allowed  to recover the scaling laws of Zimm theories (see also Ref. \cite{Larson2013}).
More recently, this method has been used for a model of solutes with several sites \cite{Poblete2014}, 
in order to mimic hydrodynamic interactions at the surface of a colloidal particle. Despite these successes, some of us 
found in a previous work that the diffusion coefficients of simple electrolytes predicted by MPCD-CC differed from those obtained by Brownian Dynamics with hydrodynamic interactions, 
in contrast with more expensive coupling algorithms \cite{Batot2013}. Indeed, one limitation of the MPCD-CC scheme is that the solute influences the solvent at the length scale of the collision cell $a_0$ only, so that 
 the hydrodynamic size of the solute is of the order of $a_0$ in any case. 
 In real systems, the hydrodynamic radius of a simple ion or of a nanoparticle is close to its structural radius derived from the atomic structure. 
It means that the structural size should also be of the order of $a_0$, or in other word that the size of the cell in MPCD should be chosen to match the structural size when the collisional coupling is used. 
 In practice, in several papers, authors suggest to choose a value of the structural diameter of the order of $a_0$ 
\cite{Hecht05,Lettinga2010,Gompper2014}, 
but there is no quantification of the effect of the cell size on the transport properties.

In the present article, we focus on the MPCD-CC method and investigate its ability 
to reproduce hydrodynamic interactions as well as transport theories at the Oseen level. 
First, we compute from different methods the effective hydrodynamic radius at infinite dilution of a solute in collisional coupling with the MPCD solvent. We show that it is in any case of the order of one third of the cell size $a_0$. If the structural radius of the solute is chosen equal to the hydrodynamic radius, it means that several solutes may be in the same collision cell at a given step, which might lead to spurious effects. We thus also investigate the influence of the cell size on the diffusion coefficient of solutes. The results are compared to the ones we previously obtained for the same systems without hydrodynamic interactions and also with MPCD and a central force between solvent and solute particles. 
For hard spheres, we find an upper limit for the cell size compared to the structural radius over which the diffusion coefficient spuriously decreases. Finally, we consider the case of solutes with an attractive interaction, as it  may increase the probability to find two solutes in the same collision cell. More precisely, we study the transport properties of simple electrolyte solutions. We compare the diffusion coefficient of ions in a 1-1 electrolyte computed by MPCD-CC  for two different resolutions of the grid to our previous numerical results, taken here as references.
We also compute the electrical conductivity of these solutions because (i) the electrical conductivity of electrolytes is known to be strongly affected by hydrodynamic couplings, (ii) a reliable semi-analytical transport theory accounting for hydrodynamic couplings at the Oseen level exists, able to predict the electrical conductivity of 1-1 electrolytes over a wide range of concentration \cite{bernard1,Dufreche05}. We find an excellent agreement between the electrical conductivity computed from MPCD-CC and the transport theory, which shows that MPCD-CC is able to capture the hydrodynamic
interactions between monovalent ions. 
However, we also show that the constraints on the choice of parameters in the MPCD-CC method prevent us from representing a real system. Finally, we use the MPCD-CC to predict the electrical conductivity of 2-1 and 2-2 electrolytes. For such systems, electrolyte transport theories are less used. Indeed, to account for the experimental conductivity, it is often necessary to use unrealistic input parameters, or to add parameters in the model, such as association constants \cite{turqcondasso}. 
Our MPCD-CC simulations shed light on the cause of this difficulty. 
Indeed, we find important differences between the simulation results and the predictions of the transport theory  for 2-1 and 2-2 electrolytes, even if the description of equilibrium properties coincide. 

The paper is organized as follows. In Section \ref{simu} we shortly describe the simulation methods and the semi-analytical theory used to compute the electrical conductivity. We compute the effective hydrodynamic radius of a solute in collisional coupling with a MPCD solvent in Sec. \ref{effective}. Then, the influence on diffusion coefficients of the size of the collision cell compared to the structural radius of the solute is investigated in Sec. \ref{grid} for solutions of neutral solutes and for electrolyte solutions. Finally, we compare the electrical conductivity
computed by MPCD-CC to the theoretical prediction in Sec. \ref{conducti}. The paper ends with a conclusive discussion. 
\section{Methods }
\label{simu}
	\subsection{Multi-particle collision dynamics simulations }
		
The fluid in MPCD is represented by pointlike particles, whose positions and velocities evolve in two steps \cite{PaddingPRE06}.  In the  streaming step, positions and velocities are propagated by integrating Newton's equations of motion. Without external forces, this yields a ballistic motion for each fluid particle $i$:
	\begin{equation}
		\mathbf{r}_{i}(t + \delta{}t_{c}) = \mathbf{r}_{i}(t) + \mathbf{v}_{i}(t)\delta{}t_{c}
		\label{streamfluid}
	\end{equation}
where $\mathbf{r}_{i}$, $\mathbf{v}_{i}$ are respectively the position and the velocity of particle $i$, and $\delta{}t_{c}$ is the time step. 
A second step, the  collision step, enables local momentum exchanges between the fluid particles. The simulation box is partitioned into cubic cells of given size $a_0$. A randomly oriented axis is defined for each collision cell, and the velocities of fluid particles relative to the velocity of the center of mass of the cell are rotated by an angle $\alpha$ around this axis:
	\begin{equation}
		\mathbf{v}_{i}(t + \delta{}t_{c}) = \mathbf{v}^{cell}_{c.o.m}(t) + \mathcal{R}_{\alpha}[\mathbf{v}_{i}(t)-\mathbf{v}^{cell}_{c.o.m}(t)]
		\label{collisionRule}
	\end{equation}
where $\mathcal{R}_{\alpha}$ is the rotation matrix and $\mathbf{v}^{cell}_{c.o.m}$ the velocity of the center of mass of the cell. The angle $\alpha$ is a fixed parameter. A random shift of the collision grid is performed at each collision step to ensure galilean invariance\cite{GalInvSRD,GalInvarianceSRD}.

The transport properties of the fluid depend on a few parameters: The number of solvent particles per cell $\gamma$,  the rotation angle $\alpha$, the time between two collisions $\delta t_c$\cite{PaddingPRE06,WinklerRev}.
It is convenient to use the fluid particle mass ${m_f}$ as the mass unit, the size of the collision cells $a_0$ as the length unit, and $k_BT$ as the energy unit with $T$ the temperature and $k_B$ Boltzmann constant. The time unit is then
\begin{equation}
t_0 = a_0\sqrt{\dfrac{m_f}{k_BT}}.
\end{equation}
The kinematic viscosity of the pure MPCD fluid is $\nu =\nu_{coll}+\nu_{kin}$
\cite{Kikuchi2003JCP,Ihle2005PRE} with
\begin{eqnarray}
	\nu_{coll} &= &\frac{1}{\lambda}\frac{(1-{\rm cos}\alpha)}{18} \left(1-\frac{1}{\gamma}+\frac{e^{-\gamma}}{\gamma} \right) \\
	\nu_{kin} &=& \lambda \left[ \frac{1}{[4-2{\rm cos}\alpha-2{\rm cos}(2\alpha)]}\frac{5\gamma}{(\gamma-1+e^{-\gamma})}-\frac{1}{2} \right].
\label{nuKinEq}
\end{eqnarray}
where $\lambda$ is the mean free path of fluid particles, $\lambda=\delta t_c/t_0$. 

By default, in what follows, the parameters of the MPCD simulations  are: $\{\alpha=130^{\circ},\gamma=5,\delta t_c=0.1t_0\}$. This ensures that the Schmidt number (ratio of the time scale of diffusive mass transfer over the time scale of momentum transfer in the fluid), corresponds to a liquid-like behavior. For this choice of parameters, the kinematic viscosity of the  fluid is $\nu=0.809 \;a_0^{2}.t_0^{-1}$ (with $\nu_{coll}  >>\nu_{kin}$), so that the dynamic viscosity $\eta$ is equal to $4.045\;m_fa_0^{-1}.t_0^{-1}$.

 Solute particles can be immersed in this solvent bath. Their dynamics is then coupled to that of the fluid particles. In what follows, we focus on the so-called collisional coupling scheme, where
solute particles i) interact with each other through a given force field, ii) participate to the collision step with solvent particles 
located in the same cell. We refer to this method as MPCD-CC hereafter. Solutes usually have a mass greater than that of fluid particles.
During the streaming step, the position $\mathbf{R}_j$ and velocity $\mathbf{V}_j$  of solute $j$ are propagated with the velocity Verlet algorithm often used in standard Molecular Dynamics (MD) simulations:

\begin{equation}
	\mathbf{R}_{j}(t+\delta{}t_{MD}) = \mathbf{R}_{j}(t) + \mathbf{V}_{j}(t)\delta{}t_{MD} + \dfrac{\mathbf{F}_{j}(t)}{2M} \delta{}t_{MD}^2, 
	\label{vvR}
\end{equation}
\begin{equation}
	\mathbf{V}_{j}(t+\delta{}t_{MD}) = \mathbf{V}_{j}(t) + \dfrac{\mathbf{F}_{j}(t)+\mathbf{F}_{j}(t+\delta{}t_{MD})}{2M} \delta{}t_{MD},
	\label{vvV}
\end{equation}
where $M$ is the mass of the solute particle, $\mathbf{F}_{j}$ is the force acting on solute $j$ at the beginning of the step, and $\delta{}t_{MD}$ is the time step. 
During the collision step, the velocities of fluid and solute particles are updated following Eq. (\ref{collisionRule}) in each collision cell. More details about this simulation scheme can be found in Refs. \cite{RipollPRE05,WinklerRev}.

Another possible coupling between the dynamics of solutes and the solvent bath consists in adding  an explicit short ranged interaction between solutes and fluid particles, preventing fluid particles to penetrate into solutes. In this case, solutes do not participate to the collision step. Details on this simulation scheme,  called MPCD-CFC (CFC for Central Force Coupling) in what follows can be found in Ref. \cite{PaddingPRE06}. In the present paper, the results obtained with MPCD-CC are compared in a few cases with those obtained in a previous study with MPCD-CFC. For these MPCD-CFC simulations, a purely repulsive short-ranged interaction potential between solutes and solvent was used \cite{Batot2013}.

	\subsection{Computation of transport coefficients and of the hydrodynamic radius from MPCD-CC}
	\label{method-transport}
The self-diffusion coefficient of solutes $D_s$ is computed using equilibrium trajectories. The mean-squared displacement as function of time is computed, and the diffusion coefficient of solutes is deduced from the slope at long time \cite{understand}:
\begin{equation}
D_s= \lim_{t\to \infty}\frac{1}{6t}\langle|{\bf R}_j(t+t_0) - {\bf R}_j(t_0)|^2\rangle_{t_0,j}  \label{msd-BD}.
\end{equation}
At infinite dilution, the hydrodynamic radius $a_{hyd}$ can be estimated  from the self-diffusion coefficient using the Stokes law, written here in the case of stick boundary conditions between solute and solvent:
\begin{equation}
 D_s=\frac{k_BT}{6\pi\eta a_{hyd}},
 \label{stokes}
\end{equation}
with $\eta$ the dynamic viscosity of the fluid, $\eta=\nu\gamma/a_0^3$. 

The electrical conductivity $\sigma$ of a solution of charged solutes is computed from equilibrium MPCD simulations using Kubo's formula \cite{understand}:
\begin{equation}
	\sigma = \frac{1}{3k_BTV} \int_0^{\infty} {\rm d}t \, 
			\langle \sum_{i=1}^{N_{ed} }q_i{\mathbf V}_i(t_0) \cdot \sum_{j=1}^{N_{ed}} q_j{\mathbf V}_j(t_0+t)
			\rangle_{t_0},
\end{equation}
with $V$ the volume of the simulation box, $N_{ed}$ the total number of solutes, $q_i$ the charge of solute $i$ and ${\mathbf V}_i$ its velocity.

We have also used non-equilibrium MPCD simulations to compute the hydrodynamic radius of a solute in collisional coupling. In this procedure, the solute particle is fixed at the center of the simulation box, and we impose a solvent flow along the $x$-direction. To induce the flow, a velocity with Gaussian distribution centered on a given value $v_0$ is added to each solvent particle situated in the layer of the simulation box perpendicular to the $x$-direction, of thickness $a_0$, at $x=0$. We have chosen $v_0=0.02$ $a_0.t_0^{-1}$. 
At stationary state, we obtain a flow around the solute with a cylindrical symmetry.  In the approximation of laminar flow ($Re \ll 1$), the analytic solution of the Stokes equation for the fluid around an isolated sphere of radius $a_{hyd}$ fixed at the origin with stick boundary conditions is given by \cite{Hulin}
\begin{eqnarray}
\label{Stokes-cyl}
v_{r} (r,x)& = &-v_{\infty}a_{hyd} \frac { x r } {2 (r^{2} + x^{2})^{\frac{3}{2}}} \\ \nonumber
v_{\theta}&= &0 \\ \nonumber
v_{x} (r,x) &= &v_{\infty} \left[ 1 - a_{hyd} \left( \frac {1} {2 \sqrt{r^{2} +
x^{2}}} + \frac{x^{2}}{2\left( r^{2} + x^{2} \right)^{\frac{3}{2}}} \right)
\right] 
\end{eqnarray}
 where cylindrical coordinates $(r,\theta,x)$ are used, $x$ being the direction of the flow and $r$ the distance to the origin.
The velocity field hence depends on two parameters: the hydrodynamic radius $a_{hyd}$ of the sphere and the fluid velocity far from the sphere $v_{\infty}$. The incompressibility of the fluid imposes $v_{\infty}$ to be equal to the average velocity within a slice of fluid in the plane perpendicular to the flow direction $x$, whatever the value of $x$.
The value of $v_{\infty}$ in the flow simulated by MPCD can thus be determined. Then, we have fitted the value of $a_{hyd}$ by minimizing the mean square deviation between the Stokes velocity field (Eq.(\ref{Stokes-cyl})) and the velocity field computed by MPCD. As the simulation box has a finite size, the simulated flow field could actually be affected by periodic boundary conditions. Nevertheless, this effect is to the first order proportional to $a_{hyd}/L_{box}$ with $L_{box}$ the size of the simulation box. As we proceed to show, under the conditions of our simulations, $a_{hyd}/L_{box}$ is less than 0.01, so that the finite size effect can be safely neglected.  

	\subsection{Theoretical computation of the electrical conductivity}
	\label{method-HNC}
    The electrical conductivity of electrolyte solutions can be predicted from analytical or semi-analytical theories related to the Fuoss-Onsager theory developed at the beginning of the twentieth century\cite{onsa57}.  These theories are based on the assumption that two main effects influence the dynamics of ions in solution, namely electrostatic interactions and hydrodynamic couplings. The forces induce modifications of the velocity of solutes which depend on the structural organisation of the solution through the direct correlation function between two solutes. Analytical expressions of the electrical conductivity have been proposed for the primitive model of electrolytes using the mean spherical approximation as closure relation to solve the Ornstein-Zernike integral equation \cite{hansen} and to compute the structural correlation functions \cite{bernard1}. In the framework of the primitive model, ions are charged hard spheres embedded in a continuous solvent of given viscosity and dielectric constant.  We propose here  to use this theoretical framework to predict the electrical conductivity of solutions of charged species of given structural and hydrodynamic radii. These theoretical results will be used as references to compare our simulation results with.

More precisely, the velocity of ion $i$ at stationary state under the presence of an electric field ${\bf E}$ reads
\begin{equation}
{\bf v}_i = \frac{D_s^0}{k_BT} q_i{\bf E} + \delta {\bf v}_i^{hyd} + \delta{\bf v}_i^{elec}, \label{tri2}
\end{equation}
where  $D_s^0$ is the self-diffusion coefficient of species $i$ at infinite dilution, related to its hydrodynamic radius through the Stokes relation (eq. (\ref{stokes})), $\delta{\bf v}_i^{hyd}$ is the hydrodynamic velocity correction due to hydrodynamic interactions, and $\delta{\bf v}_i^{elec}$ is the velocity correction due to electrostatic couplings. 
The hydrodynamic velocity correction  $\delta{\bf{v}}_i^{hyd}$ can be approximated using the Oseen tensor as \cite{Dufreche05}:
\begin{equation}
\delta {\bf v}_i^{hyd}=\sum_j n_j q_j {\bf E}  \frac{2}{3\eta} \int_0^{\infty}rh_{ij}(r)\mathrm{d}r \label{hydro_int} 
\end{equation}
where $n_j$ is the density of species $i$ and $h_{ij}(r)$ is the total pair correlation function between species $i$ and $j$ at a distance $r$. 
The velocity correction due to electrostatic couplings, $\delta{\bf v}_i^{elec}$ also depends on the equilibrium pair correlation function, and scales with $1/\eta$. The full expression of this term can be found {\em e. g.} in \cite{bernard1,serge}. Here, we use the HyperNetted Chain closure relation \cite{hansen} to compute the total pair correlation function between ions $h_{ij}(r)$. We have checked that for the systems investigated here these correlation functions were very close to those obtained by MPCD. The advantage of the HNC solution for the correlation function is the absence of statistical noise. Moreover, the hydrodynamic radius $a_{hyd}$ and the structural radius $a_{HS}$ are two independent parameters of the calculation, $a_{hyd}$ being related to $D_s^{\circ}$ and $a_{HS}$ being a parameter of the interaction potential. 
This will allow us to use exactly the same parameters as those of the MPCD-CC simulations to compute the electrical conductivity. 

This theory has been challenged against experimental data for a large variety of 1-1 electrolytes. The unique parameter that can be adjusted is the radius of the ions, the infinite dilution diffusion coefficient being directly extrapolated from low concentration data. The calculated electrical conductivity was found in excellent agreement with experimental data up to high concentrations \cite{bernard1,DurandvidalJPC96,Dufreche05,RogerJPCB2009} for several electrolytes, either using the crystallographic radius\cite{bernard1}, or using a radius deduced from other quantities, such as the osmotic coefficients \cite{Dufreche05}. Nevertheless, for asymmetric electrolytes, {\em e. g.} 2-1 or 2-2 electrolytes, the theory accounts for experiments only if the radius of the divalent ion is artificially small\cite{bernard1}, or with an additionnal fitting parameter, such as an association constant between ions \cite{turqcondasso}.

\section{Effective hydrodynamic radius of a solute in the MPCD-CC scheme}
\label{effective}
\begin{figure}[h!]
\centering
\includegraphics[scale=0.37]{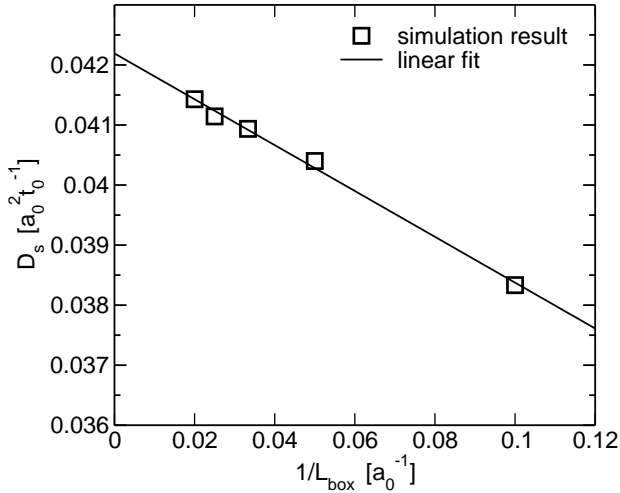}
\caption{Self-diffusion coefficient of MPCD-CC solutes at infinite dilution as a function of the inverse box length. There are no direct interactions between solutes. The linear fit of the data writes $D_s=0.0422 - 0.0382 \frac{1}{L_{box}}$. }
\label{fig1}
\end{figure}

The effective hydrodynamic radius of a solute can be deduced from its self-diffusion coefficient at infinite dilution from the Stokes law. To compute the diffusion coefficient of a solute in collisional coupling with the MPCD solvent at infinite dilution, we performed simulations at low density of solutes ($N_{ed}/L_{box}^3$ between $0.0008$ for the largest simulation box, and $0.01$ for the smallest one, with $N_{ed}$ the number of solutes), without any
 direct interactions between solutes (the solute-solute correlations are those of a perfect gas). 
Such simulations require very long trajectories to get enough statistics.
The simulations were run for embedded particles of mass $M=10m_f$, for several box lengths between $10$ $a_0$ and $50$ $a_0$.  Seven independent trajectories of $2.5\times10^7$ steps each, with $\delta t_{MD}=0.01$ $t_0$ were run. The results are shown in Fig. \ref{fig1}.

We observe in Fig. \ref{fig1} that the self-diffusion coefficient is a linear function of the inverse of the box length due to finite size effects   \cite{hasi,Yeh2004JPCB}.  This scaling was first predicted by Hasimoto \cite{hasi} who studied the velocity flow of a fluid across a periodic cubic array of spheres with $L_{box}$ being the distance between two spheres. For a fluid of viscosity $\eta$ we expect
\begin{eqnarray}
D_s(L_{box})&=&D_s(\infty)-2.837\frac{k_BT}{6\pi\eta L_{box}}\\
&=& D_s(\infty)-\frac{0.0372}{ L_{box}},
\label{hasimoto}
\end{eqnarray}
with $\eta=4.045\;m_f\,a_0^{-1}\,t_0^{-1}$ and $k_BT=1$. This prediction is in excellent agreement with the fit of our data. 
It shows that our simulations do correspond to infinite dilution conditions, with a very weak effect of solute-fluid interactions on the fluid viscosity, and hence no visible hydrodynamic interactions between solutes inside the simulation box. 
This also enables us to extrapolate our results to infinite box size so that we obtain
$D_s(\infty)=D_s^\circ=0.0422\;a_0^2t_0^{-1}$, and the corresponding hydrodynamic radius of the solute from the Stokes law (eq.(\ref{stokes})): $a_{hyd}=0.31\;a_0$.  This value is very close to the radius obtained by other authors in a MPCD solvent with the parameters $\{\alpha=130^{\circ},\gamma=10,\delta t_c=0.1t_0\}$ for a solute of mass $M=10m_f$.  They found $D_s^\circ=0.02\;a_0^2t_0^{-1}$ so that $a_{hyd}=0.3\;a_0$\cite{SinghJCP2014,Poblete2014}. From our value of the hydrodynamic radius, we can compute an effective packing fraction for the systems investigated here. It is always lower than $4. 10^{-4}$, which confirms that we are in the high dilution regime for indirect hydrodynamic interactions between solutes in the simulation box.  

\begin{figure}[h!]
\centering
\includegraphics[scale=0.35]{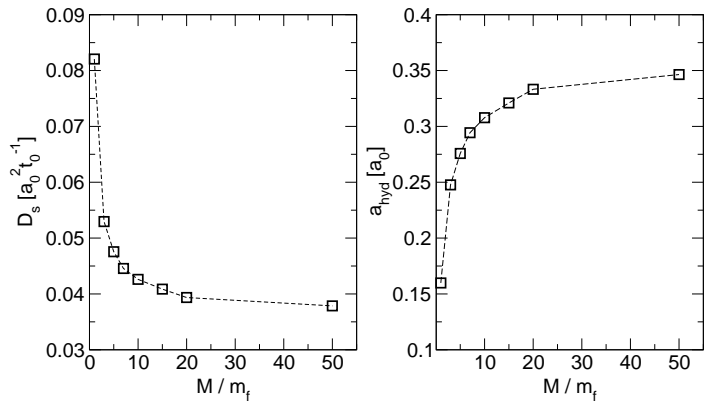}
\caption{Left: Self-diffusion coefficient at infinite dilution of MPCD-CC solutes as functions of the mass of solutes, for a dilute system with no direct interactions ($N_{ed}/L_{box}^3=0.0013$). Right: hydrodynamic radius $a_{hyd}$ deduced from  $D_s$ using the Stokes relation (Eq.(\ref{stokes})).  }
\label{figDvsM}
\end{figure}

The hydrodynamic radius of a solute in collisional coupling deduced from the diffusion coefficient at infinite dilution is thus found smaller than half the size of the collision cell. One may wonder if we could vary the effective hydrodynamic radius by changing the mass of the solute, which would modify the momentum exchange with the solvent particles during the collision step. We show in Fig.\ref{figDvsM}-left the self-diffusion coefficient at infinite dilution as a function of the mass $M$ of the solute. Note that the results have been extrapolated to a simulation box of infinite size using eq.(\ref{hasimoto}).  We observe that the diffusion coefficient is an increasing function of the mass of the solute and reaches a plateau for large values of $M/m_f$, more precisely for $M$ larger than $\approx 2\gamma$. This behavior was already observed by Ripoll {\em et al} \cite{RipollPRE05}. Indeed, the heavier a solute is, the more it slows down solvent particles that participate in the collision step in the same cell. From these results, we  extract the effective hydrodynamic radius from Stokes law (see Fig.\ref{figDvsM}-right). It appears then that the hydrodynamic radius can not exceed $0.35a_0$. We have also computed the self-diffusion at infinite dilution for a solute of mass $M=5m_f$ in a MPCD solvent with $\gamma=10$, keeping the same values of $\alpha$ and of $\delta t_c$. We have obtained $D_s^\circ=0.02595\;a_0^2t_0^{-1}$. For these parameters,  the viscosity of the solvent is $\eta=8.7\;m_fa_0^{-1}.t_0^{-1}$, so that the effective hydrodynamic radius in this case is $a_{hyd}=0.23\;a_0$. If we assume that the influence of the solute mass on the hydrodynamic radius is almost the same as the one described in Fig. \ref{figDvsM}, the case  $M=5\;m_f, \,\gamma=10$ should be close to the case $M=3\;m_f, \,\gamma=5$. Indeed, we  obtained $a_{hyd}\approx0.25\;a_0$ for $M=3$ $m_f, \,\gamma=5$. Therefore, increasing the density of the MPCD fluid should not allow us to increase the hydrodynamic radius significantly.  

\begin{figure}[ht!]
\begin{center}
\includegraphics[scale=0.37]{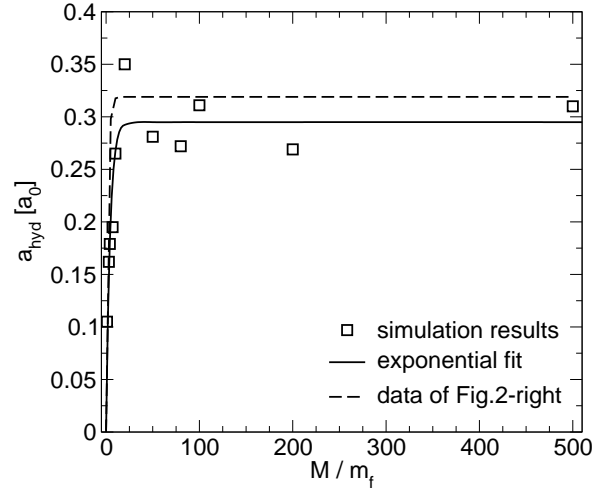}
\caption{Hydrodynamic radius $a_{hyd}$ of a solute in collisional coupling with a MPCD solvent determined from non-equilibrium simulations as a function of the solute mass $M/m_f$. The solid line is an exponential fit of the data. The dashed line corresponds to an exponential fit of the data given in Fig. \ref{figDvsM}-right. }
\label{radiusCC}
\end{center}
\end{figure}

Another route to determine the  hydrodynamic radius of a solute is to induce a solvent flow around the solute from non-equilibrium simulations, and to compare the simulated flow to the analytical result, as described in Sec. \ref{method-transport}. We have checked that the results were independent from the value of the solvent velocity in the small velocity regime.
We have again studied the influence of the solute mass on its hydrodynamic radius.
The mass of the particles varied from $M =  m_{f}$ to $M = 1000$ $m_{f}$. The Stokes flow field around a sphere with stick boundary conditions was well reproduced. The values of the hydrodynamic radius were fitted in order to minimize the difference between the computed and analytical velocity flow. We give in Fig. \ref{radiusCC} the obtained hydrodynamics radii. In every case the root mean squared errors between computed and analytical velocities were between 1\% and 2.5\%. 
The results again show that the hydrodynamic radius is an increasing function of the mass which can be fitted by an exponential function, $a_{hyd}(M)=0.295*(1-exp(-M/3.3))$. When the solute fixed at the center of the simulation box is heavy enough, the solvent particles in its cell are basically stopped because of the collision step, so that the hydrodynamic radius becomes independent from the solute mass at large mass. We also represent in Fig. \ref{radiusCC} with a dashed line the exponential fit of the data obtained from equilibrium simulations (data of Fig. \ref{figDvsM}-right). Both methods, equilibrium and non-equilibrium simulations yield the same limiting value of the hydrodynamic radius at large solute mass within the statistical uncertainty: $a_{hyd}\approx0.3$~a$_{0}$.
	
The fact that the hydrodynamic radius of solutes in MPCD-CC is in any case of  the order of $0.3\;a_0$ is clearly a limitation of the method.  For a given structural model of the solute (for instance the radius for a hard sphere model of solutes), the resolution of the MPCD grid relative to the size of the particles is imposed if one aims at keeping the structural radius close to the hydrodynamic one. Moreover, if the resolution of the grid is not thin enough, two or more solute particles can be located in the same MPCD collision cell at the same time. This raises the question of the validity of the method in such cases: What is the impact of the pathological description of short ranged hydrodynamic interactions when they emerge from momentum exchange including several solutes particles ?
In the literature, the resolution of the grid is chosen so that the size of the collision cells roughly equals the diameter of the solute particles. This minimizes the probability that several solute particles participate in the collision step in the same cell. However, such choice leads to another kind of issue: The hydrodynamic radius is then different from the structural one. 
For a solute hard sphere diameter equal to $1.2\;a_0$, the structural radius is indeed twice larger than the hydrodynamic one. This may lead to a decrease of hydrodynamic interactions, as we proceed to show. 

\section{Influence of the size of the collision cell on the diffusion coefficient}
\label{grid}
We study in this part the influence of the resolution of the collision grid on the diffusion coefficient of solutes in collisional coupling with the solvent. We expect spurious effects on the dynamics of solutes when several solutes participate to the collision step in the same cell. These effects should be larger when the size of the cell increases and exceeds  the minimal distance of approach between solutes, and/or when the concentration of solutes increases, and/or when attractions between solutes exist. 
	\subsection{Solutes modelled as hard spheres}
\begin{figure}[ht!]
\begin{center}
\includegraphics[scale=0.37]{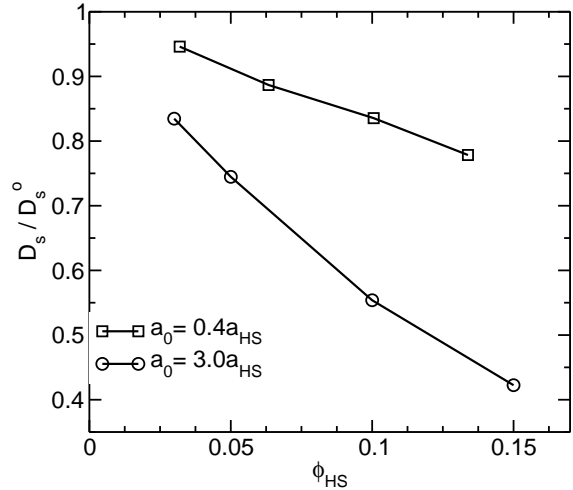}
\caption{Diffusion coefficient of neutral hard spheres as function of their packing fraction $\phi_{HS}$, using MPCD-CC simulations. The size of the collision cells is either $a_0=0.4\;a_{HS}$, or $a_0=3.0\;a_{HS}$. }
\label{DCC}
\end{center}
\end{figure}

We have used the hard-sphere molecular dynamics algorithm \cite{Wainwright59} to compute the trajectories of solute particles between two collision steps in MPCD-CC. First, the influence of the volume fraction on the self-diffusion coefficient was studied for two different values of $a_0$: $a_0=0.4\;a_{HS}$ and $a_0=3.0\;a_{HS}$ with $a_{HS}$ the hard sphere radius of the solute. For $a_0=0.4\;a_{HS}$ the structural size is thus very close to the hydrodynamic one. The mass of the solute is $M=10\;m_f$ in every case. 
We present in Fig. \ref{DCC} the computed self-diffusion coefficients divided by the value at infinite dilution ($D_s^{\circ} = 4.22$ $10^{-2}\,a_0^2 t_0^{-1}$) as functions of the volume fraction $\Phi_{HS}=(4/3\pi N_{ed})(a_{HS}/L_{box})^3$. We have checked that the results obtained using a short-ranged Week-Chandler-Anderson interaction potential and a standard molecular dynamics algorithm instead of the hard-sphere algorithm coincide exactly. 

As expected, the diffusion coefficient is in every case a decreasing function of the solute density (see Fig. \ref{DCC}). However, the influence of the size of the collision cell relative to that of the solute particles is striking.
First, in the case $a_0=3.0\;a_{HS}$, the diffusion coefficients are considerably smaller than with $a_0=0.4\;a_{HS}$. Second, the decrease of the diffusion coefficient with the volume fraction is much more pronounced with $a_0=3.0\;a_{HS}$. This choice of the resolution of the grid  is clearly not correct. We have thus more precisely investigated the influence of the size of the collision cell on the diffusion coefficient.

\begin{figure}[ht!]
\begin{center}
\includegraphics[scale=0.37]{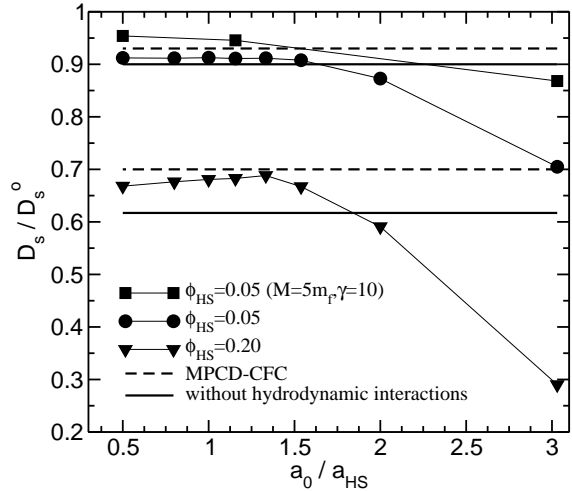}
\caption{Diffusion coefficient of neutral hard spheres as function of the size of the collision cell $a_0$ relative to the radius of the solute particle $a_{HS}$, using MPCD-CC simulations. Two packing fractions are considered, $\phi_{HS}=0.05$ and a crowded case $\phi_{HS}=0.20$. The reference dashed lines correspond to MPCD with central force coupling, and the solid ones to Brownian Dynamics without hydrodynamic interactions. }
\label{DCC2}
\end{center}
\end{figure}

We give in Fig.\ref{DCC2} the diffusion coefficients obtained at a given volume fraction as functions of the ratio between the size of the collision cell and the hard-sphere radius of solute $a_0/a_{HS}$. 
We focus on two volume fractions, $\phi_{HS}=0.05$  and $\phi_{HS}=0.20$.
For small values of $a_0$, {\em i.e.} for thin collision grids, the diffusion coefficient is almost independent from the value of $a_0$ (see the circles and the triangles symbols in Fig. \ref{DCC2}). 
However, the diffusion coefficient of the solute starts to decrease dramatically for $a_0$ larger than $1.5\;a_{HS}$. At the highest volume fraction, 
the value of $D_s/D_s^\circ$ drops by $60\%$  when $a_0$ increases from $1.4\;a_{HS}$ to $3.0\;a_{HS}$. This behavior is less pronounced in the dilute case, because the probability to have several solutes in the same collision cell is smaller:  $D/D^\circ$ drops by $22\%$ when $a_0$ increases from $1.4\;a_{HS}$ to $3.0\;a_{HS}$. This strong decrease of the diffusion coefficient comes from the spurious coupling between solutes when they are in the same collision cell. 
Indeed, the maximum distance between two solutes in a collision cell is equal to $\sqrt3 a_0\approx1.73\;a_0$, so that
we must have $2\;a_{HS}>1.73\;a_0$ or $a_0/a_{HS}<1.16$ to ensure that two hard-sphere solutes cannot be in the same collision cell. When $a_0$ is larger than this value, it may happen that two solutes participate in the collision step in the same cell.
For a few values of $a_0/a_{HS}$  in the dilute case, we have changed the ratio between the solute mass and the average density of the MPCD solvent in a cell ($M=5\;m_f$ with $\gamma=10$). In this case, the infinite dilution diffusion coefficient is $D_s^\circ=0.02595\;a_0^2t_0^{-1}$. We expected the problem of momentum exchange between solutes to be reduced in this case: If two solutes are in the same collision cell, their total mass stays of the order of that of the solvent also participating to the collision. Nevertheless, if the decrease of the diffusion coefficient when $a_0/a_{HS}$ increases is smaller in these conditions, it is still clearly visible (square symbols in Fig. \ref{DCC2}).

To get more insight into these results, we compared the value of the diffusion coefficient with references obtained in a previous study from our group \cite{Batot2013}. In this previous study, we compared the predictions of Brownian Dynamics (BD) simulations with hydrodynamic interactions at the Rotne-Prager level, BD without hydrodynamic interactions, and MPCD simulations with a central force coupling (CFC) between solutes and a MPCD fluid. Self-diffusion coefficients obtained from BD with hydrodynamics and from MPCD-CFC were in excellent agreement for neutral solutes interacting through a WCA potential. These results are independent from the size of the collision cell, and are represented in Fig. \ref{DCC2} as horizontal dashed  lines for volume fractions equal to $0.05$ and $0.20$. We also report in the same figure the results obtained without hydrodynamic interactions as horizontal solid lines. 

For the less concentrated system, $\phi_{HS}=0.05$, the solid and dashed lines are close to each other because hydrodynamic interactions have a weak influence on the diffusion coefficient. The MPCD-CC results stand between the reference results with and without hydrodynamic interactions, which means that  this method is unable to capture such subtle effect. 

In the more concentrated case, hydrodynamic interactions have a non-negligible effect on the diffusion coefficient: $D_s/D_s^\circ$ is equal to $0.61$ without HI and to $0.70$ with HI. In this case, we observe a rather good agreement between the diffusion coefficient predicted by MPCD-CC and the reference for small  $a_0$ values. 
$D_s/D_s^\circ$ computed by MPCD-CC is actually found to weakly increase when $a_0$ increases, at small $a_0$ values. This is due to the increase of hydrodynamic interactions because of the increase of $a_{hyd}$ in this range. Indeed, as shown in previous section, $a_{hyd}\approx 0.3\;a_0$. In the limit of small $a_0$, the hydrodynamic radius is thus negligible compared to the structural radius $a_{HS}$.
At  $a_0/a_{HS}=0.5$ we have $a_{hyd}\approx0.15\;a_{HS}$ and at $a_0/a_{HS}=1.3$, we have $a_{hyd}\approx0.39\;a_{HS}$. This increase of $a_{hyd}$ leads to an increase of the hydrodynamic interactions. $D_s/D_s^\circ$ reaches a maximum at about $a_0/a_{HS}=1.3$ but  for larger $a_0$ values several solutes can interact via the collision step. 
It should be noted that the maximum value of $D_s/D_s^\circ$ computed by MPCD-CC is in excellent agreement with our reference. 
On the contrary, for larger values of $a_0$,  the MPCD-CC results tend towards those obtained without HI and  becomes even smaller than them.

It is interesting to look at what happens for $a_0/a_{HS}=2$, which is the most common choice in the literature when MPCD-CC is used \cite{Hecht05,Lettinga2010,Gompper2014}.
For both volume fractions of $0.05$ and $0.2$, the value of the diffusion coefficient is below the one that was computed without hydrodynamic interactions. Does this mean that there are no hydrodynamic interactions ? No, since for this kind of parameters one can recover some features of hydrodynamic interactions, such as the scaling of polymer diffusion coefficients with polymer size. However, one should be careful when interpreting the results because an artificial coupling between solutes located in the same cell can happen in this range.

	\subsection{Solutes with electrostatic interactions}

\begin{figure}[ht!]
\begin{center}
\includegraphics[scale=0.37]{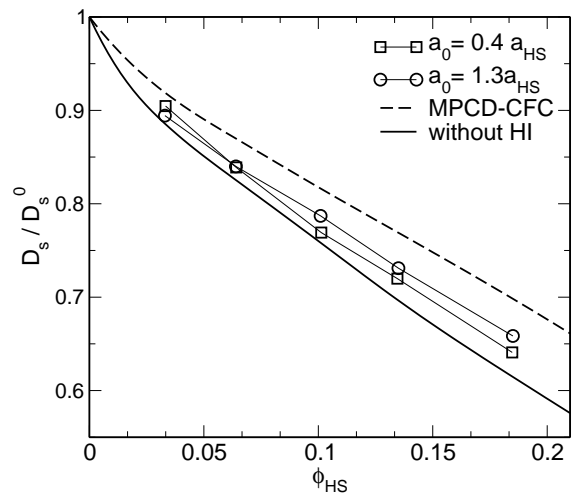}
\caption{Self-diffusion coefficient of charged hard spheres (1-1 electrolyte) as function of their volume fraction $\phi_{HS}$.
The size of the collision cell is either $a_0=0.4\;a_{HS}$, or $a_0=1.3\;a_{HS}$.
The reference dashed line corresponds to MPCD with central force coupling, and the solid one to Brownian Dynamics without hydrodynamic interactions.
}
\label{Dsel}
\end{center}
\end{figure}

One may wonder whether the attraction between oppositely charged particles can change the impact of the size of the collision cell on the transport coefficients of solutes. 
To answer this question, we have computed the self-diffusion coefficients of a mixture of charged hard spheres of opposite charges, 
for several volume fractions between $0.033$ and $0.185$. 
The system corresponds to a  1-1 electrolyte solution described by the primitive model of electrolytes. 
Ions are hard-spheres of charge equal to $+1$ and $-1$ in a dielectric continuum of relative permittivity $\varepsilon_r$ . Electrostatic interactions between ions are computed thanks to an Ewald summation \cite{understand} with a conductive boundary condition. 
 The characteristic electrostatic length scale, the Bjerrum length $l_b={e^2}/{(4\pi\varepsilon_0\varepsilon_r k_BT)}$ with $e$ the elementary charge, $\varepsilon_0$ the permittivity of vacuum and $\varepsilon_r$ the relative permittivity of the solvent, is equal to $0.71$ nm, 
corresponding to water at room temperature. The structural radius of ions is $a_{HS}=l_b/3.57$. 
We have compared the results for two different values of the cell size, $a_0/a_{HS}=\{0.4; 1.3\}$. For these grid resolutions, we have shown in the previous section that no spurious effect on the diffusion coefficient of neutral hard spheres was observed. 
As previously, we compare the self-diffusion coefficients to the reference values obtained in our previous paper i) with hydrodynamic interactions (MPCD with Central Force Coupling), and ii) without hydrodynamic interactions (Brownian Dynamics without HI). 

Results are shown in Fig. \ref{Dsel}. They are close to those obtained with neutral hard-spheres: Diffusion coefficients increase slightly when the size of the collision cell increases, because the hydrodynamic radius increases, and this effect is more pronounced at large volume fractions. However, the computed values are not closer to
the reference with hydrodynamic interactions than to the reference without hydrodynamic interactions. This  
contrasts with the case of neutral hard-spheres for $\phi=0.2$ and $a_0/a_{HS}=1.3$ for which the computed diffusion coefficient was clearly closer to the reference with hydrodynamic interactions. 
The interplay between electrostatic and hydrodynamic effects may explain this discrepancy, since attractive coulombic interactions increase
the impact of short range hydrodynamic interactions.

In the following, our goal is to compare MPCD-CC simulations with another reference that is known to describe accurately hydrodynamic interactions,  with a ratio $a_{hyd}/a_{HS}$ that is exactly the same as with MPCD-CC.   
In what follows, we have chosen to compute another transport coefficient of electrolyte solutions using MPCD-CC, the electrical conductivity, because a reliable semi-analytical theory of this quantity is available, which includes hydrodynamic interactions quantitatively.

\section{Electrical conductivity of an ionic solution}
\label{conducti}
The measurement of the electrical conductivity is a widely used technique to analyze an electrolyte solution or a charged colloidal suspension known to be affected by hydrodynamic interactions. As ions of opposite charge move and drag the solvent in opposite directions under an electric field, hydrodynamic interactions between ions strongly reduce the conductivity compared to the infinite dilution value. 
Within the framework of the semi-analytical theory described in Section \ref{method-HNC}, the hydrodynamic radius and the structural radius of ions can be treated as independent parameters. The structural radius  $a_{HS}$ is a parameter of the interaction potential within the primitive model of electrolytes and is involved in the integral equations solved with the HNC closure to compute the pair distribution functions. 
Moreover, this radius is used as the lower bound of the integral which allows us to compute the electrostatic relaxation velocity correction  $\delta {\bf v}_i^{elec}$.
The effective hydrodynamic radius of the charged solute which is related to the infinite dilution diffusion coefficient $D_s^\circ$ through the Stokes law influences i) the electrical conductivity at infinite dilution, also called the ideal conductivity, ii) the electrostatic relaxation velocity correction  $\delta {\bf v}_i^{elec}$. 
When this theoretical framework is used to predict the electrical conductivity of real systems,  $a_{HS}$ is usually obtained by mapping the equilibrium properties of the system to that of the primitive model of electrolytes, for example by fitting the 
 osmotic
 coefficients \cite{Dufreche02,Dufreche05}. 
The hydrodynamic size $a_{hyd}$ is usually deduced from the experimental asymptotic self-diffusion coefficient at infinite dilution  \cite{DurandVidalJPC06}. For most 1-1 electrolytes, $a_{HS}$ and $a_{hyd}$ are close to each other \cite{Dufreche02,Dufreche05}.

In our case, the comparison between MPCD-CC simulations and the theory will enable us to check whether both methods predict the same influence of hydrodynamic and electrostatic interactions on the electrical conductivity, for a given value of the hydrodynamic radius constrained by the MPCD-CC technique. In what follows, we first study the same model of 1-1 electrolyte solution as in the previous section, with the size of the collision cell  $a_0/a_{HS}=1.3$. We stick to this value as it maximises the intensity of hydrodynamic interactions, as shown previously, without leading to spurious effects as two solutes cannot be in the same collision cell. 
Then, we investigate the case of electrolytes with divalent ions, namely 2-1 and 2-2 electrolytes. Both ions of the 2-2 electrolytes have the same radius, again with $a_0/a_{HS}=1.3$. For the 2-1 electrolyte, we study two different models: in the first one, named model A, the divalent cation and the monovalent anion have the same radius ($a_0/a_{HS}=1.3$). In the second model, named model B, the cation is larger than the anion:  $a_0/a_{cation}=0.93$ and $a_0/a_{anion}=1.3$. Model B is more realistic than model A, as simple cations like the calcium ion are usually larger than simple monovalent anions like the chloride ion (we are considering here hydrated ions).

MPCD-CC simulations were run for $2$ $10^{5}$ $t_0$. If the value of $t_0$ in real units is obtained by mapping the viscosity of water, it corresponds to a physical time of $23$ ns. The electrolyte concentration was varied in MPCD-CC by changing the number of ions in the simulation box, keeping in every case the same box size. 
We have checked in every case that the radial distribution functions between ions obtained from MPCD-CC simulations were very close to those obtained within the HNC closure for every distance within the simulation box. The interest to use HNC total pair distribution functions in the theory instead of those computed from simulations is that they are obtained for larger distances and do converge to zero without statistical noise.
 For systems where there is a small discrepancy between pair distribution functions obtained by HNC and by MPCD-CC, we have checked that the electrical conductivity computed from the semi-analytical theory was the same in both cases. 
In what follows, we give the electrical conductivity $\sigma$ divided by its value at infinite dilution $\sigma^\circ$, which reads for solutions of charged solutes

\begin{equation}
\sigma^\circ= \frac{ e^2 E (z_{+}^2N_{+}+z_-^2N_-) D_s^\circ}{k_BTL_{box}^3} 
\end{equation}
with $e$ the elementary charge, $E$ is the value of the electric field, $N_+$ (resp. $N_-$) the number of positive (resp. negative) ions in the simulation box, and $z_+$ (resp. $z_-$) is the valency of the positive (resp. negative) ions. 

\subsection{MPCD-CC is able to predict the electrical conductivity of a 1-1 electrolytes}
\begin{figure}[ht!]
\begin{center}
\includegraphics[scale=0.37]{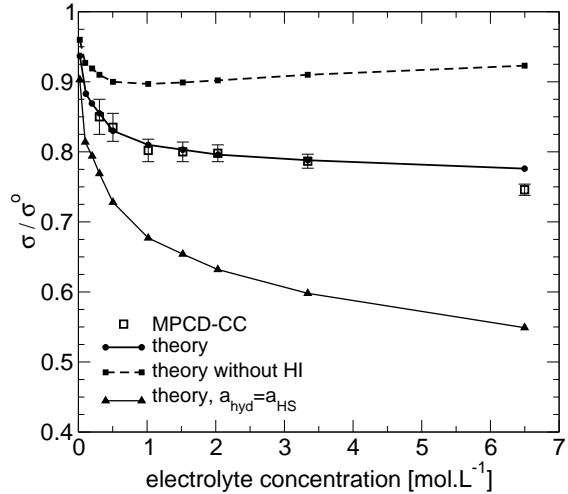}
\caption{Electrical conductivity $\sigma$ divided by the value at infinite dilution $\sigma^{\circ}$, as a function of the electrolyte concentration.  Open squares: MPCD-CC simulations with $a_0/a_{HS}=1.3$; Filled circles: HNC-transport theory with the same parameters as MPCD-CC ($D_s^{\circ} = 4.22$ $10^{-2}\,a_0^2 t_0^{-1}$, $a_{HS}=a_0/1.3$); Filled squares: HNC-transport theory without hydrodynamic interactions; Filled triangles: HNC-transport theory with $a_{hyd}=a_{HS}$.
}
\label{comp_cond}
\end{center}
\end{figure}

We give in Fig. \ref{comp_cond} the electrical conductivity as a function of the molar concentration of the 1-1 electrolyte $c=N_{+}{\cal{N}}_A/(L_{box}^3)$ with ${\cal N}_A$ the Avogadro number. Error bars of the MPCD-CC results (open squares) are derived from the comparison of 10 independent simulations for each concentration. Results are compared to those obtained with the transport theory (filled circles).

The agreement between MPCD-CC and the theoretical predictions is excellent. 
For all systems but the more concentrated one, 
the values derived from the theory largely fall within the error bars of MPCD-CC results. 
For the system at $6.3$ mol.L$^{-1}$, it is impossible to say whether the disagreement comes from a limitation of the theory or from the simulations.      
We also plot in Fig. \ref{comp_cond} the results obtained with the theory without hydrodynamic interactions (filled squares and dashed lines): Only the correction to the velocity due to the electrostatic couplings, $\delta {\bf v}_i^{elec}$, was taken into account in the theory. 
It is equivalent to suppress hydrodynamic interactions from the transport theory. We see in Fig.  \ref{comp_cond} that
the values of the electrical conductivity are much higher in this case, and closer to the ideal conductivity ($\sigma/\sigma^\circ$ is close to $1$). Moreover, we plot the theoretical results obtained by using $a_{hyd}=a_{HS}$ instead of $a_{hyd}=0.4\,a_{HS}$ (see the triangles in Fig. \ref{comp_cond}), i.e. in a case where hydrodynamic couplings are expected to be increased. The electrical conductivity is found to be strongly decreased compared to the case where $a_{hyd}=0.4\,a_{HS}$. This illustrates the fact that MPCD-CC simulations describe hydrodynamic interactions that are significantly weaker than those of real systems, for which the hydrodynamic radius is very close to the structural one. This is a limitation to have in mind when comparing MPCD-CC results to experimental data.
In conclusion, these results  show that : (i)  MPCD-CC is able to capture hydrodynamic couplings corresponding to an hydrodynamic radius $a_{hyd}=0.4\,a_{HS}$, and (ii) the agreement between the theory with hydrodynamic couplings 
and the MPCD-CC requires that both treatments include hydrodynamic couplings with a quantitatively similar intensity.
 
 \subsection{MPCD-CC reveals the limitations of the semi-analytical transport theory for 2-1 and 2-2 electrolytes}
 
\begin{figure}[ht!]
\begin{center}
\includegraphics[scale=0.37]{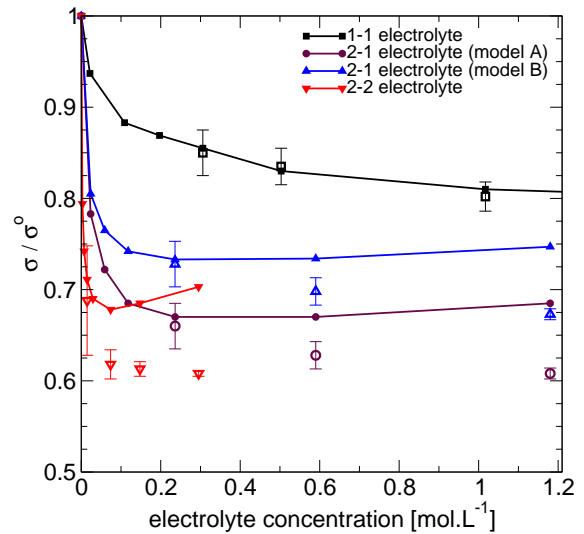}
\caption{Electrical conductivity $\sigma$ divided by the value at infinite dilution $\sigma^{\circ}$, as a function of the electrolyte concentration for different electrolytes.  Open symbols: MPCD-CC simulations; Filled symbols and lines : HNC-transport theory with the same parameters as MPCD-CC. }
\label{cond2-1}
\end{center}
\end{figure}

We give in Fig. \ref{cond2-1} the electrical conductivity of 2-1 and 2-2 electrolytes as a function of the molar concentration of the electrolyte $c=N_{+}{\cal{N}}_A/(L_{box}^3)$ with ${\cal N}_A$ the Avogadro number. The electrical conductivity of the 1-1 electrolyte is also shown in this figure. 
For the 2-1 electrolytes, we observe on Fig. \ref{cond2-1}  that the electrical conductivity of model B is larger than that of model A for all concentrations, both in simulation and theory. The only difference between these two models is the size of the cation, which is larger in model B than in model A. Actually, increasing the structural radius leads to a decrease of electrostatic and hydrodynamic couplings between ions, resulting in a lower conductivity. 

We observe for the 2-1 and 2-2 electrolytes a large discrepancy between MPCD-CC results and the prediction of the transport theory, except for the most dilute cases. 
Semi-analytical calculations overestimate the conductivity of the solution. This result gives an explanation to the use of unrealistically small radii as input parameters of the theory in previous studies of 2-1 and 2-2 electrolytes. For instance, in \cite{bernard1}, the structural radius of the magnesium cation is taken equal to $1.3$~\AA, which is very small for an hydrated divalent ion. More generally, the overestimation of the conductivity by the theoretical calculation when multivalent ions are present 
is often attributed to defaults in the description of the equilibrium pair distribution functions of ions, and not to dynamical ingredients of the theory. 
These equilibrium defaults either come from a default of the model (the primitive model does not explicitely account for ion polarisability for instance), or from the approximations within the integral equation theory, which implies that the correct structure of the primitive model is not obtained by solving Ornstein-Zernike equation with a given closure equation. In both cases, the long range ion-ion correlation are usually correctely described, but short range interactions need to be added. The simplest way to add these contact effects is to define an effective equilibrium constant between ions \cite{turqcondasso}. The primitive model is still used, but with the presence of new species, the ion pairs.

One of the major interests of simulations is to separate the effects of the model {\em per se} from the influence of the approximations of the theory. Here, we use the primitive model in both MPCD simulations and HNC theory. Therefore, pair formation due to subtile hydration effects or dispersion interactions are not described in any cases. The combined use of simulations and theory allows us to show that the difference between MPCD results and predictions from the electrolyte transport theory has nothing to do with the equilibrium ingredients within the theory. The ion-ion pair distribution functions from HNC and those from the simulations are very close to each other. If we take the pair distribution functions from the MPCD simulations as input of the transport theory, we find very small differences (less than $1 \%$) compared to the transport theory fed with HNC equilibrium pair distributions. An important dynamical effect seems to be missing in the semi-analytical electrolyte transport theory. Some additionnal work is needed to characterize in more details the nature of the differences in the dynamical treatment.  
 
\section{Conclusion}
The use of mesoscopic simulation techniques to interpret dynamical properties of solutions usually requires special care, in particular 
when one is interested in a quantitative mapping between a real system and the simulated model. 
Traditional techniques, such as Brownian Dynamics, benefit from the fact that they are tightly related to the usual theories, 
and that the input parameters such as the infinite dilution diffusion coefficient or the hydrodynamic radius are 
widely used. Experimental studies have been improved to determine this class of parameters. More recent mesoscopic techniques, such 
as MPCD, rely on input parameters that are totally specific to the technique, and can be related to the properties of the fluid 
within a particular set of approximations. In order to understand the limitations of such methodologies, it can be useful to 
quantify commonly used dynamical quantities such as the hydrodynamic radius.   
    
In this study, we find that the effective radius of a solute in collisional coupling with a MPCD solvent is about one third the size of the collision cell. 
To ensure that two hard-sphere solutes cannot be in the same collision cell, we should have in principle $a_0/a_{HS}<1.16$ with $a_{HS}$ the hard-sphere radius of the solute. Moreover, we have shown that for neutral hard-spheres, hydrodynamic interactions between solutes are  maximum at about $a_0/a_{HS}=1.3$. In addition to these results, it should be noted that since the mean free path $\lambda$ needs to stay close to $0.1$ to keep the Schmidt number low \cite{RipollPRE05}, the value of the kinematic viscosity $\nu$
is also constrained. Such constraints on the choice of MPCD parameters can be rephrased in term of time scales. 
When the solution is infinitely diluted, there are no hydrodynamic interactions. The only time scale of importance for transport is given by the infinite dilution diffusion coefficient, or by the product $\eta a_{hyd}$. The MPCD dynamics can be mapped onto a real time dynamics by mapping the MPCD diffusion coefficient to the value of the real diffusion coefficient of the solute.     
When solute concentration increases, a second important time scale emerges, the time scale at which hydrodynamic interaction are propagated, through the diffusion of momentum in the solution. This time scale is related to the kinematic viscosity $\nu$. 
These two time scales are intimately related, as the diffusion coefficient depends on viscosity:
At infinite dilution, we have $D_s^\circ=k_BT/(6\pi\eta a_{hyd})$. 
The constraint on the hydrodynamic radius $a_{hyd}$ which depends on the size of the collision cell makes it impossible to have both $D$ and $\nu$ match the values of a real system. 
For instance, let us consider a typical ion in water at ambient temperature, like a chloride or a sodium ion. Its diffusion coefficient is of the order of  $10^{-9}\;m^2s^{-1}$, and the dynamic viscosity of water is close to $10^{-3}\;kg\,m^{-1}s^{-1}$. The structural radius of an ion is about $0.15\,nm$. 
In our MPCD simulation, for typical values of the parameters ($M=10$ $m_f$, $\gamma=5$, $\alpha=130$, $\lambda=0.1$), 
the diffusion coefficient is equal to $4.22$ $10^{-2}$ $a_0^2 t_0^{-1}$, and the kinematic viscosity is equal to $0.809$ $a_0^2 t_0^{-1}$. 
The value of the MPCD time unit $t_0$ can be mapped into a real unit by choosing either the diffusion coefficient $D$, setting $D_{MPCD}=D_{real}$ or $\eta$, setting $\nu_{MPCD}=\nu_{real}$. 
The two values of $t_0$ obtained by these mappings differ by one order of magnitude. 

In the case of a polymer made of $N$ monomers of infinite dilution diffusion coefficient $D_s^\circ$, within the Zimm theory, one can express the diffusion coefficient as a sum of a term depending 
on the diffusion coefficient, and a term depending directly on the viscosity and on an equilibrium property (the average inverse distance between monomers):

\begin{equation}
D=\frac{D_s^\circ}{N}+\frac{k_BT}{6\pi\eta}\langle \frac{1}{r_{ij}}\rangle
\end{equation}
The typical scaling appears when the first term becomes negligible, {\em i. e.} 
when the infinite dilution diffusion time scale is no longer relevant, and only hydrodynamic interactions affect transport.  
In such case, the main dynamic properties of the system should be qualitatively reproduced, and the MPCD time scale can be mapped onto 
the dynamics of the real system by setting $\nu_{MPCD}=\nu_{real}$. This is in agreement with the results obtained with polymers using the same simulation methodology. Although the hydrodynamic friction per monomer is very small, the correct scaling of diffusion properties on the number of monomers per polymer chain recovers the Zimm prediction, with $\langle \frac{1}{r_{ij}}\rangle$ computed using the adequate theory of polymer equilibrium properties. 

On the other hand, things might become more complicated when the scaling regime is not reached, for instance when finite size effects are important, typically for small values of $N$ in the case of polymers. In the case of electrolytes, for most systems there is no regime in which hydrodynamic interactions totally dominate. In such case, although it is not possible to quantitatively map the dynamics of MPCD to that of a real system, the mesoscopic simulation can be used to assess the validity of a theory for which the parameters have been artificially chosen to be equal to those in MPCD. Such strategy is indeed the only possible way to explore the range of validity of theories for systems such as suspensions of nanoparticles or nanoporous media. Indeed, for such systems the interpretation of experiments often relies on concepts and parameters, such as the zeta potential, whose relationship with the underlying molecular system can be questionable \cite{Wesolowski2016,Siboulet2017}.      
We made in the present article an attempt to quantify the limitations of a common transport theory of electrolytes. This theory belongs to the family of treatments of hydrodynamic and electrostatic interactions presented in the seminal work of Fuoss and Onsager. These theories have mainly been extended to account for more sophisticated treatments of the equilibrium correlation between the electrolyte species, but the original dynamical treatment has not evolved much (see {\em e.g.} \cite{Pusset2015}).  We found that there are strong differences between the theory and the simulations for 2-1 and 2-2 electrolytes that are not due to the description of the equilibrium properties of the system. An important dynamical effect seems to be missing in the semi-analytical electrolyte transport theory. Some additionnal work is needed to characterize in more details the nature of the differences in the dynamical treatment. In a future article, we will extend this study to suspensions of small charged nanoparticles, and use MPCD simulation as a tool to test the approximations of the transport theories used in these systems.

\section{Acknowledgments}
Partial financial support of the Agence Nationale de la Recherche in the
frame of the project Celadyct (ANR-12-BS08-0017-01) is gratefully
acknowledged. 
\newpage

\bibliographystyle{apsrev4-1}
\bibliography{Biblio-avril-2018-propre}
\end{document}